\documentclass[12pt,titlepage]{article}
\usepackage{amsmath}
\usepackage{amssymb}
\usepackage{graphicx}
\usepackage{caption2}
\usepackage{amsfonts}
\usepackage{cite}

\newcommand{\be}{\begin{equation}}
\newcommand{\ee}{\end{equation}}
\newcommand{\bea}{\begin{eqnarray}}
\newcommand{\eea}{\end{eqnarray}}
\newcommand{\bef}{\begin{figure}}
\newcommand{\ef}{\end{figure}}
\newcommand{\bt}{\begin{tabular}}
\newcommand{\et}{\end{tabular}}
\newcommand{\bno}{\begin{enumerate}}
\newcommand{\eno}{\end{enumerate}}

\def\3{\ss}

\catcode`\"=\active
\def"{\accent'177}
\begin{document}

\begin{center}
{\large\bf  On the  numerical simulation of propagation of micro-level inherent uncertainty for chaotic dynamic systems}

Shijun  Liao $^{a,b,c}$

$^a$ Department of Mathematics,  $^b$ State Key Lab of Ocean Engineering\\
$^c$  School of Naval Architecture, Ocean and Civil Engineering\\
Shanghai Jiao Tong University, Shanghai 200240, China\\
(email: sjliao@sjtu.edu.cn)

\end{center}

\hspace{-0.75cm}{\bf Abstract}   In this paper, an extremely accurate  numerical algorithm, namely the ``clean numerical simulation'' (CNS), is proposed to accurately simulate the propagation of micro-level inherent physical uncertainty of chaotic dynamic systems.   The chaotic Hamiltonian H\'{e}non-Heiles system for motion of stars orbiting in a plane about the galactic center is used as an example to show its basic ideas and validity.   Based on  Taylor expansion at rather high-order and MP (multiple precision) data in very high accuracy,  the CNS approach can  provide  reliable trajectories of the chaotic system  in a finite interval $t\in[0,T_c]$, together with an explicit estimation of the critical time $T_c$.   Besides, the residual and round-off errors are   verified and estimated  carefully  by means of different time-step $\Delta t$,  different   precision of data,  and  different  order $M$ of Taylor expansion.  In this way, the numerical  noises  of  the  CNS  can be reduced to a required level, i.e.  the CNS is a rigorous algorithm.  
It is illustrated that, for the considered problem,  the truncation and round-off errors of the CNS can be  reduced even to the level of $10^{-1244}$ and $10^{-1000}$, respectively,  so that the micro-level inherent physical uncertainty of the initial condition (in the level of $10^{-60}$) of the H\'{e}non-Heiles system can be investigated accurately.    It is found  that, due to the sensitive dependence  on initial condition (SDIC) of chaos,  the micro-level inherent physical uncertainty of the position and velocity of a star transfers into the macroscopic randomness of motion.   Thus,  chaos  might be  a bridge from the micro-level  inherent  physical uncertainty to  the macroscopic randomness in nature.    This might provide us a new explanation to the SDIC of chaos from the physical viewpoint.    

\hspace{-0.75cm}{\bf Key Words} Chaos;  Clean Numerical Simulation;  Micro-level uncertainty

\section{Introduction}

Using high performance digit computers,  a lots of complicated problems in science, finance and engineering have been solved with satisfied accuracy.  However,  there exist  some  problems which are still rather difficult to solve even by means of the most advanced  computers.  One of them is  the propagation of micro-level  inherent  {\em physical} uncertainty  of  chaotic  dynamical  systems.   

It is well-known that all numerical simulations are not ``clean'':  there exist more or less numerical noises such as truncation  and round-off errors, which greatly depend on numerical algorithms.    In most cases,  such kind of numerical noises are much larger than the micro-level   inherent  physical  uncertainty of dynamic systems under consideration, so that the micro-level  inherent  uncertainty is completely lost in the numerical noise.  This becomes more serious for chaotic dynamic systems, which have the sensitive dependence on initial conditions (SDIC), i.e. very tiny  change  of initial condition leads to great difference of numerical simulations of chaotic systems so that  long-term prediction is impossible.   Thus,  very  fine  numerical  algorithms  need be developed   to  accurately simulate  the  propagation   of micro-level inherent physical uncertainty of chaotic dynamic systems.  This is the motivation of this article.

In this article,  a kind of numerical algorithm in a rather high accuracy, called the ``clean numerical simulation'' (CNS), is proposed to accurately simulate propagation of micro-level inherent physical uncertainty of chaotic dynamic systems.   Here, the word ``clean'' means that the truncation  and  round-off  errors  can be controlled  to  an  arbitrary  level  that is much less than the micro-level  inherent  physical uncertainty of the initial condition so that the numerical noises can be neglected in a  given finite  interval  of  time  for the propagation of  uncertainty.       A chaotic Hamiltonian system  proposed  by   H\'{e}non and Heiles \cite{Henon1964} is used  to show its validity.     The basic ideas of the so-called clean numerical simulation (CNS) are given in \S 2, followed by the investigation of the micro-level uncertainty of the system in \S 3 and its propagation in \S 4 from statistical  viewpoint.    Conclusions and discussions are given in \S 5.

\section{The numerical algorithm of the CNS}
  
 \subsection{Basic ideas}
  
H\'{e}non and Heiles \cite{Henon1964}  proposed a Hamiltonian system of equations 
\begin{eqnarray}
\ddot{x}(t)  &=& -x(t) -2 x(t)   \; y(t) ,    \label{geq:1} \\
\ddot{y}(t)  &=&  -y(t) -x^2(t) + y^2(t),  \label{geq:2}
\end{eqnarray} 
to approximate the motion of stars orbiting in a plane about the galactic center, where the dot denotes the differentiation with respect to the time $t$.   Its solution is chaotic for some initial conditions, such as 
\begin{equation}
 x(0)=\frac{14}{25}, y(0)=0, \dot{x}(0)=0, \dot{y}(0)=0, \label{ic:case1}
 \end{equation}
  as mentioned by Sprott \cite{Sprott2010}.     Without loss of generality,  let us use this chaotic system to describe the basic ideas of the CNS and to illustrate  its validity.   

It is well-known \cite{Egolf2000, Gaspard1998, Li1975, Lorenz1963, Lorenz1993, Smith1998,  Sprott2010, Werndl2009}  that chaotic dynamic systems have the sensitive dependence on initial conditions (SDIC), i.e.  a tiny change  of initial conditions leads to great difference of numerical simulations at large time, so that long-term prediction of chaos is impossible.   It is well-known that all numerical simulations contain the unavoidable truncation and round-off errors at each time-step.   Generally speaking,   most of traditional numerical simulations of chaos are mixed with these numerical noises and thus are not ``clean''.    Because  these  numerical noises of traditional numerical approaches are generally much larger than the micro-level inherent physical  uncertainty of initial condition,    the propagation of such kind of physical uncertainty of chaotic dynamic systems  has  never  been studied accurately, to the best of the author's knowledge.   

For numerical simulations of  chaotic dynamic system,  we must take rigorous account of numerical errors and rounding, because ``what is  observed on the computer  screen would be completely unrelated to what was meant to be simulated'', as pointed out by Galatolo et al \cite{GHR2012}.   The methods of shadowing may gain accurate numerical simulations closed to true trajectories of hyperbolic dynamic systems, but fail  to have long shadowing  trajectories  for those with a fluctuating number of positive finite-time Lyapunov exponents,  as pointed out by Dawson et al \cite{Dawson1994}.   Besides, it is found that numerical simulations of chaotic systems given by  low-order   Runge-Kutta methods or Taylor expansion approaches  have sensitive dependence not only on initial conditions but also on numerical algorithms, so that  different numerical schemes might lead to completely different  long-term predictions, as pointed out by Lorenz \cite{Lorenz1989, Lorenz2006}  and Teixeira et al.  \cite{Teixeira2007} .    

 In order  to gain reliable chaotic solutions in a long  interval of time,  Liao \cite{Liao2009}  developed a numerical technique with  extremely high accuracy, called  here  the ``clean numerical simulation'' (CNS).   Using  the computer algebra system Mathematica  with the 400th-order Taylor expansion for continuous functions and data in accuracy of 800-digit precision,  Liao \cite{Liao2009}  gained, for the first time,  the  reliable  numerical results of chaotic solution of Lorenz equation  in a long interval $0\leq t \leq 1175$ LTU (Lorenz time unit).   The basic ideas of the CNS are simple and straightforward.    
 Since  the order of  Taylor expansion is very high,  the corresponding  truncation  error is rather small.  Besides,  since   all data are expressed in the accuracy of large-number digit precision,  the small enough round-off error is guaranteed.    Thus,  as long as the order of Taylor expansion is high enough and the digit-number of data is long enough,  both of the truncation and round-off errors can be  much smaller than  the  micro-level  inherent physical  uncertainty so that the propagation of micro-level  uncertainty  of the initial condition  can be simulated accurately in a long enough interval of time.   Here, the ``clean'' numerical simulation means that the truncation and round-off errors can be controlled to an arbitrary  level  so that the numerical noises can be neglected in a  given  {\em finite}  interval  of  time, as shown later.      Currently, Liao's ``clean'' chaotic solution\cite{Liao2009} of Lorenz equation is confirmed by Wang et al \cite{Wang2011} to be a reliable trajectory of Lorenz equation in the interval $0 \leq t \leq 1175$ LTU,  who  used  parallel computation with the multiple precision (MP) library:   they gained reliable chaotic solution of Lorenz equation up to 2500 LTU by means of  the 1000th-order Taylor expansion and  data in the accuracy of 2100-digit precision.      Note that, similar to the so-called shadowing trajectories given by the shadowing approach \cite{Sauer1997},  such kind of ``clean'' numerical simulations given by the CNS  are close to true trajectories  of  chaotic systems.  

The  CNS is based on  Taylor expansion at a rather high-order.     Let $(x_n, y_n)$ and $(\dot{x}_n, \dot{y}_n)$ denote the position and velocity  at the time $t_n = n \Delta t$, where $\Delta t$ is a constant time-step.   Assume that  $x(t), y(t)$ are $M+1$  times differentiable on the open interval $(t, t+\Delta t)$ and continuous on the closed interval $[t, t+\Delta t]$.  According to Taylor theorem,  we have 
\begin{eqnarray}
x(t+\Delta t)  & =&  x(t) +  \sum_{n=1}^{M} a_n(t) \; (\Delta t)^n + R_M^x(t), \\ 
 y(t+\Delta t) &= &  y(t) + \sum_{n=1}^{+\infty} b_n(t) \; ( \Delta t)^n + R_M^y(t), 
\end{eqnarray} 
where  
\begin{equation}
a_n(t) =\frac{1}{n!} \frac{d^n x(t)}{d t^n}= \frac{x^{(n)}(t)}{n!},  \;\;\;  b_n(t)  = \frac{1}{n!} \frac{d^n y(t)}{d t^n} = \frac{y^{(n)}(t)}{n!}
\end{equation}
and
\begin{eqnarray}
R^x_M(t) &=& a_{M+1}(\xi_1) \; (\Delta t)^{M+1} , \hspace{0.5cm}  t \leq \xi_1 \leq t +\Delta t,\\
R^y_M(t) &=& b_{M+1}(\xi_2) \; (\Delta t)^{M+1} , \hspace{0.5cm}  t \leq \xi_2 \leq t +\Delta t,
\end{eqnarray}
are remainders of $x(t)$ and $y(t)$, respectively.   Assuming that 
\begin{equation}
|a_{M+1}(t)| <  \mu , \;\; |b_{M+1}(t)| < \mu, \;\;\; t > 0,  \label{bounded}
\end{equation}
it holds obviously 
\begin{equation}
\left|   R_M^x(t)  \right| <  \mu \; (\Delta t)^{M+1}, \;\; \left|   R_M^y(t)  \right| < \mu \; (\Delta t)^{M+1}.  \label{Reminder}
\end{equation}
Thus,  we have the following theorem

\hspace{-0.75cm} {\bf Theorem of truncation error}  {\em If $x(t), y(t)$ are $M+1$  times differentiable on the open interval $(t, t+\Delta t)$ and continuous on the closed interval $[t, t+\Delta t]$, and besides if $|x^{(M+1)}(t)| /(M+1)! < \mu$ and $|y^{(M+1)}(t)| /(M+1)! < \mu$ for $t > 0$, where $\mu > 0$ is a constant, then the Taylor expansion}
\begin{eqnarray}
x(t+\Delta t)  &\approx &  x(t) +  \sum_{n=1}^{M} a_n \; (\Delta t)^n,   \label{x:Taylor}\\ 
 y(t+\Delta t) &\approx &  y(t) + \sum_{n=1}^{+\infty} b_n \; ( \Delta t)^n,  \label{y:Taylor}
\end{eqnarray} 
{\em have the truncation errors less than}  $\mu \; (\Delta t)^{M+1}$.

The round-off error is  determined by the accuracy of data.   To avoid large round-off error,   all data are expressed in high accuracy of long-digit  precision.   For example, one can use data in accuracy of $2M$-digit  precision, where $M$  is the order of Taylor expansions  (\ref{x:Taylor}) and   (\ref{y:Taylor}).   Thus, for large enough $M$, the round-off error are rather small.  For example, in case of $M=70$, all data are expressed  in accuracy of 140-digit  precision so that the corresponding round-off error is in the level of $10^{-140}$.    Such kind of high precision data can be gained easily  by means of computer algebra system like Mathematica and Maple, or the multiple precision (MP)  library for FORTRAN and C.   Obviously, the larger the value of $M$, the smaller the truncation  and the round-off errors.  In this meaning,  we can control the truncation  and round-off errors  to a  required level.   

The coefficients $a_n$ and $b_n$ can be calculated in  a  recursive way.   Assume that  $a_0=x_n, b_0=y_n, a_1=\dot{x}_n, b_1=\dot{y}_n$ are known.  Substituting the Taylor expansions (\ref{x:Taylor}) and (\ref{y:Taylor})  into the original governing equations (\ref{geq:1}) and (\ref{geq:2}) of the  H\'{e}non and Heiles   system  \cite{Henon1964}  and equaling the like power of $\Delta t = t-t_n$, we have the recursion formula
\begin{eqnarray}
a_{n+2} &=& -\frac{a_n +2\sum\limits_{k=0}^n a_k b_{n-k}}{(n+1)(n+2)},\\
b_{n+2} &=& -\frac{b_n +\sum\limits_{k=0}^n\left(a_k a_{n-k}-b_k b_{n-k} \right) }{(n+1)(n+2)}
\end{eqnarray}
for $n\geq 0$.  Then, we have the $M$th-order Taylor approximation 
\begin{eqnarray}
x_{n+1} \approx \sum_{k=0}^{M} a_k \;  (\Delta t)^k, \;\;  y_{n+1} \approx \sum_{k=0}^{M} b_k \;  (\Delta t)^k
\end{eqnarray}
and
\begin{eqnarray}
\dot{x}_{n+1} &\approx &\sum_{k=0}^{M-1} (k+1) a_{k+1} \;  (\Delta t)^k, \\
 \dot{y}_{n+1} &\approx & \sum_{k=0}^{M-1} (k+1) b_{k+1} \;  (\Delta t)^k
\end{eqnarray}
at the time $t_{n+1}= (n+1)\Delta t$.   Besides,   all data are expressed here  in the accuracy of $2M$-digit precision (we use the computer algebra system Mathematica).        In this way,  one gains  rather  accurate  numerical simulations of $x(t)$ and $y(t)$ step by step in a finite interval of time,   with extremely  small  truncation and round-off errors at each time-step, as  verified  below.   

For short time,  both of the truncation and round-off errors  are so small that the numerical results are often close to the true trajectory.  This is the reason why most of numerical results of chaotic systems given by different approaches  match well in a short time from the beginning.   It is widely believed by the scientific community that such kind of numerical results of chaos in a short time is reliable.    However, due to the sensitivity on initial conditions of chaotic dynamic system,   the truncation  and round-off errors are amplified  quickly  so that the numerical results depart greatly from the true trajectory after a critical  time $T_c$.    Here,  $T_c$ denotes  such a  maximum  time that  numerical  results gained by means of different numerical approaches (for example, with different  $M$ and $\Delta t$ of the CNS)  are close to the true trajectory of chaotic solution in the interval $0 \leq t \leq T_c$.   In other words, the numerical results are ``clean'', i.e.  without observable influence by the round-off and truncation errors,  and thus is reliable in the finite interval $t\in[0,T_c]$.    Here,  the  so-called critical predictable time $T_c$ is  similar  to  the so-called shadowing time for the shadowing approach \cite{Dawson1994, Sauer1997}.   Mathematically, let $u_1(t)$  and $u_2(t)$ denote two time-series given by different numerical  approaches.  The so-called  ``critical  time'' $T_c$  is determined by the criteria of decoupling 
\begin{equation}
\left|  1- \frac{u_1}{u_2}\right|> \delta, \;\;\;  \dot{u}_1 \; \dot{u}_2 < -\epsilon, \hspace{0.5cm} \mbox{at  $t = T_c$}, \label{Tc:criteria}
\end{equation}
where $\epsilon > 0$ and $\delta > 0$ are two small constants ($\epsilon =
1$  and $\delta = 5\% $ are used in this article).   In this paper, the critical  time $T_c$  is determined by the CNS approach, i.e. the values of $M$, $\Delta t$ and the accuracy  of data.  Obviously, the larger $M$, the smaller $\Delta t$ and the higher accuracy of data,  the longer time interval $[0,T_c]$ in which  the numerical results match well with the true trajectory.    For given reasonable $\Delta t$ and high accuracy of data, the larger the value of $M$, the larger $T_c$.    So,  $T_c$ for given $M$ is determined by comparing the corresponding  CNS  result with that obtained by means of a larger value of $M$ with the same initial condition, the same $\Delta t$ and the same accuracy of data.              
 
The key  step of the CNS is to provide a good  estimation  of the critical time   $T_c$,  which is  an important characteristic length-scale of time for the CNS.      Without loss of generality,  we  use  in this article the $M$th-order Taylor expansions (\ref{x:Taylor}) and (\ref{y:Taylor})  with $\Delta t = 1/10$ and the data in accuracy of $(2M)$-digit precision.  Comparing different CNS results given by different $M$,  we gain the different values of  $T_c$ for different $M$ by means of the criteria (\ref{Tc:criteria}).    Then, by means of regression analysis,   it is found that  $T_c$ can be approximately expressed by 
\begin{equation}
T_c \approx 32 (1 + M).  \label{Tc:M}
\end{equation}
 For details of how to gain the above estimation  of $T_c$,  please refer to Liao \cite{Liao2009}.   Seriously speaking,  given two time series $u_1(t)$ and $u_2(t)$,  different small values of $\epsilon$ and $\delta$  might give a little different value of $T_c$.   However,    it is found that the estimation expression of $T_c$  is not sensitive to the values of $\epsilon$ and $\delta$, mainly because chaotic systems are  sensitive to numerical noises.    Thus,   (\ref{Tc:M}) provides us a good estimation of the critical time $T_c$.   For the sake of guarantee,   it is  better to choose a little larger value of $M$ than that estimated  by  (\ref{Tc:M}) in practice.    For example,  
 in order to gain reliable chaotic solution of the H\'{e}non and Heiles system  \cite{Henon1964}  in  the interval  $0\leq t \leq 2000$, say, $T_c = 2000$,   we  use  the 70th-order\footnote{The estimation formula (\ref{Tc:M}) gives $M  \approx  62$ for $T_c = 2000$.     Considering that (\ref{Tc:M}) is an estimation formula for the chaotic Hamiltonian H\'{e}non-Heiles system,  we choose $M= 70$ so as to ensure that the  CNS  results are  indeed  reliable trajectories in the interval $0\leq t \leq 2000$. } Taylor expansion  (with $\Delta t = 1/10$)  and the data in accuracy of 140-digit precision.   It should be mentioned here that 
(\ref{Tc:M}) is consistent with the conclusion  about methods of shadowing \cite{Sauer1997}:   the shadowing time have power law dependencies on the level of numerical noise.   

 Thus, given an arbitrary value of $T_c$, we can always  calculate such a corresponding order $M$ of Taylor expansions  that the corresponding  CNS  result is reliable in the interval $t\in[0,T_c]$, as verified below.      In other words,  given  the  critical  time $T_c$,  the choice of the time-step $\Delta t$ and the order $M$ of Taylor expansion for  reliable  trajectories  in $t\in[0,T_c]$ is  under  control.     In this meanings,  the CNS approach can be   regarded as a ``rigorous'' one.                  

\subsection{Validity of numerical simulations}

\begin{table}[htdp]
\caption{Reliable numerical results of  H\'{e}non and Heiles'  chaotic  system (\ref{geq:1}), (\ref{geq:2}) and (\ref{ic:case1}) given by $ M =  70$ and $\Delta t =1/10$ with  data in accuracy of 140-digit  precision.}
\begin{center}
\begin{tabular}{|c|c|c|} \hline\hline
$t$ & $x(t)$  &  $y(t)$ \\  \hline
500 &  0.19861766  &  -0.23842431  \\
1000 &  -0.04915404 &  -0.31971648  \\
1100 &  -0.48949729  & -0.04052161 \\
1200 &  -0.04886847  &  0.77797491 \\
1300 &   0.03097135  &  0.32401254  \\
1500 &  0.03489977  &  0.43408169    \\
2000 &  0.44371428 & -0.30558921  \\
\hline\hline
\end{tabular}
\end{center}
\label{Table:case1}
\end{table}%

As mentioned before,  the larger the order $M$ of Taylor expansion and  the more accurate  the data, the better the corresponding CNS results  of chaotic system (\ref{geq:1}) and (\ref{geq:2}).      The CNS  results at $t$ = 500, 1000, 1500 and 2000 given by $M=70$ in case of the initial condition (\ref{ic:case1}) are listed in Table~\ref{Table:case1}.   

Is it a reliable  trajectory  of the chaotic system (\ref{geq:1}), (\ref{geq:2}) and (\ref{ic:case1}) in the interval $0\leq t \leq 2000$?     To  verify  its validity,  we  repeated  computations by means of $\Delta t =1/10$ and $M=100, 150, 200, 300, 500$, respectively,  and found that  {\em all} of them give exactly the {\em same}  trajectory   in the interval $0\leq t \leq 2000$,  as listed in Table\ref{Table:case1}.     Besides, even using a smaller time-setp $\Delta t =1/20$ and $\Delta t =1/100$ of the chaotic system (\ref{geq:1}), (\ref{geq:2}) and (\ref{ic:case1}), we  {\em always} gain the exactly {\em same} trajectory   in the finite interval $0\leq t \leq 2000$   by means of $M=100$, 150, 200, 300 and $500$, respectively.    All of these indicate that the CNS approach indeed provide us a {\em reliable}  trajectory of the chaotic system (\ref{geq:1}) and (\ref{geq:2}) under the initial condition (\ref{ic:case1}) in the {\em finite} interval $0\leq t \leq 2000$.

\begin{table}[htdp]
\caption{ Estimated level of the truncation and round-off errors  of the CNS results of the chaotic system (\ref{geq:1}), (\ref{geq:2}) and (\ref{ic:case1})  in case of $\Delta t = 1/10$.}
\begin{center}
\begin{tabular}{|c|c|c|c|} \hline\hline
$M$	&	constant $\mu$ for (\ref{bounded}) 	&	truncation error	&  round-off error \\
\hline 
70	&	$10^{-29}$	&	$10^{-100}$	&	10$^{-140}$ \\  
100	&	$10^{-44}$	&	$10^{-145}$	&	10$^{-200}$ \\  
150	&	$10^{-68}$	&	$10^{-219}$	&	10$^{-300}$ \\  
200	&	$10^{-93}$	&	$10^{-294}$	&	10$^{-400}$ \\  
300	&	$10^{-143}$	&	$10^{-444}$	&	10$^{-600}$ \\  
500	&	$10^{-243}$	&	$10^{-744}$	&	10$^{-1000}$ \\  
\hline\hline
\end{tabular}
\end{center}
\label{Table:error:dt=0.1}
%\end{table}%

%\begin{table}[htdp]
\caption{ Estimated level of the truncation and round-off errors  of the  CNS results of the chaotic system (\ref{geq:1}), (\ref{geq:2}) and (\ref{ic:case1})  in case of $\Delta t = 1/20$.}
\begin{center}
\begin{tabular}{|c|c|c|c|} \hline\hline
$M$	&	constant $\mu$ for (\ref{bounded}) 	&	truncation error	&  round-off error \\
\hline 
70	&	$10^{-29}$	&	$10^{-122}$	&	10$^{-140}$ \\  
100	&	$10^{-44}$	&	$10^{-176}$	&	10$^{-200}$ \\  
150	&	$10^{-68}$	&	$10^{-265}$	&	10$^{-300}$ \\  
200	&	$10^{-93}$	&	$10^{-355}$	&	10$^{-400}$ \\  
300	&	$10^{-143}$	&	$10^{-535}$	&	10$^{-600}$ \\  
500	&	$10^{-243}$	&	$10^{-895}$	&	10$^{-1000}$ \\  
\hline\hline
\end{tabular}
\end{center}
\label{Table:error:dt=0.05}
%\end{table}%

%\begin{table}[htdp]
\caption{ Estimated level of the truncation and round-off errors  of the  CNS results of the chaotic system (\ref{geq:1}), (\ref{geq:2}) and (\ref{ic:case1})  in case of $\Delta t = 1/100$.}
\begin{center}
\begin{tabular}{|c|c|c|c|} \hline\hline
$M$	&	constant $\mu$ for (\ref{bounded}) 	&	truncation error	&  round-off error \\
\hline 
70	&	$10^{-29}$	&	$10^{-170}$	&	10$^{-140}$ \\  
100	&	$10^{-44}$	&	$10^{-245}$	&	10$^{-200}$ \\  
150	&	$10^{-68}$	&	$10^{-369}$	&	10$^{-300}$ \\  
200	&	$10^{-93}$	&	$10^{-494}$	&	10$^{-400}$ \\  
300	&	$10^{-143}$	&	$10^{-744}$	&	10$^{-600}$ \\  
500	&	$10^{-243}$	&	$10^{-1244}$	&	10$^{-1000}$ \\  
\hline\hline
\end{tabular}
\end{center}
\label{Table:error:dt=0.01}
\end{table}%

To verify the CNS results, let   us further  consider the level of truncation and round-off errors.  
In case of $M=70$ and $\Delta t =1/10$, the round-off error is in the level of $10^{-140}$.   The corresponding truncation error of the CNS approach  can be roughly estimated in the following way.    According to our CNS results,   the maximum values of $|a_{70}|$ and $|b_{70}|$ are $6.1 \times 10^{-34}$ and $6.7 \times 10^{-34}$,  respectively.    Since two divergent series  decouple quickly due to the sensitive dependence  on numerical noises,  the Taylor series  should be convergent in the interval $t\in[0,T_c]$,  i.e.  
\[      \frac{|a_{71}| \Delta t}{|a_{70}|}  <1, \;\;  \frac{|b_{71}| \Delta t }{|b_{70}|}   <1.    \]
 Thus,   we have the estimation    
\[  |a_{71}| < |a_{70}|/\Delta t  <  6.1 \times 10^{-33},  \;\;\;     |b_{71}| < |b_{70}|/\Delta t < 6.7 \times 10^{-33}.   \]   
Although there exist some uncertainty in the above deduction,   we have many reasons to  assume  that\footnote{Here, we multiply the values at the right-hand side of the above expressions by $10^4$ and replace the number 6.1 and 6.9 by 1.0 for the sake of simplicity.} 
\begin{equation}   
  |a_{71}| < 10^{-29}, \;\;\;  |b_{71}| < 10^{-29},   \label{bound:M=70}    
\end{equation}
i.e. $\mu = 10^{-29}$.   Then,  according to (\ref{Reminder}),  the truncation errors should be  less than $10^{-100}$, which is  rather small.   Similarly,   the truncation errors in case of $\Delta t=1/10$ and $M=100, 150, 200, 300$ and 500 are less than $10^{-145}$, $10^{-219}$, $10^{-294}$, $10^{-444}$ and $10^{-744}$, respectively, as shown in Table~\ref{Table:error:dt=0.1}.  

Similarly, in case of $M=70$ and $\Delta t =1/20$,   the maximum CNS  results of $|a_{70}|$ and $|b_{70}|$ are $6.1 \times 10^{-34}$ and $6.9 \times 10^{-34}$,  respectively, so that  the two inequalities in  (\ref{bound:M=70}) still hold, say, we have the same constant $\mu=10^{-29}$ for (\ref{bounded}) to be valid,  although  the smaller time step $\Delta t$ is used.   It is found that,  in case of $M=70$ with much smaller time-step $\Delta t=1/100$,  the corresponding  maximum CNS values of $|a_{70}|$ and $|b_{70}|$ are $6.2 \times 10^{-34}$ and $7.1 \times 10^{-34}$, which are very close to those found in case of $M=70$ with $\Delta t = 1/10$ and $\Delta t = 1/20$,  so that   we  still have the {\em same} constant $\mu=10^{-29}$ for (\ref{bounded}) to be valid!   In fact, according to our numerical simulations based on the CNS approach, it is found that, for the same $M$ but different time-step $\Delta t\leq 1/10$,  the two inequalities in (\ref{bounded}) indeed  hold with the same constant $\mu$, as shown in Table~\ref{Table:error:dt=0.1},  Table~\ref{Table:error:dt=0.05} and Table~\ref{Table:error:dt=0.01}.   All of these verify the validation of (\ref{bounded}) and therefore the correction of our estimation for the truncation errors.           

Note that, the larger the order $M$ of Taylor expansion and the smaller the time-step $\Delta t$,  the smaller the truncation and round-off errors, as shown in Table~\ref{Table:error:dt=0.1} to Table~\ref{Table:error:dt=0.01}.  Especially, 
  in case of $M=500$ and $\Delta t = 1/100$,  the corresponding truncation error  is in the level of $10^{-1244}$ and the round-off error is in the level of $10^{-1000}$, respectively,  which are much  smaller than those given by  $M=70$ and $\Delta t = 1/10$,  so that we have many reasons to believe that the numerical result given by $M=500$ and $\Delta t=1/100$ is much closer to the true trajectory of chaotic system (\ref{geq:1}) and (\ref{geq:2}) under the given initial condition (\ref{ic:case1}).  However,  it should be emphasized that {\em all} of our CNS results given by  $M\geq 70$ and $\Delta t \leq 1/10$ are the {\em same} as those listed in Table~\ref{Table:case1}.   In other words,  the CNS provides us  the chaotic  results that are {\em independent} of not only the order $M$ of Taylor expansion but also the time-step $\Delta t$ and the data precision.   This guarantees that our CNS results  given by means of  70th-order Taylor expansion and  data in accuracy of 140-digit precision are indeed a true, reliable trajectory of the chaotic dynamic system (\ref{geq:1}) and (\ref{geq:2}) with the initial condition (\ref{ic:case1}), at least  in the interval $t\in[0,2000]$.  
  
According to Tables~\ref{Table:error:dt=0.1} to \ref{Table:error:dt=0.01}, the truncation and round-off error of the CNS approach can be decreased to the level of $10^{-1244}$ and $10^{-1000}$ (by means of $\Delta t = 1/100$ and $M=500$), respectively.   Thus, theoretically speaking,  the truncation and round-off error of the CNS approach can be reduced to a required level.   Besides,  the CNS results given by $\Delta t = 1/10$ and $M=70$  agree well (in the accuracy of 8-digit precision) with {\em all}  of the CNS results by  the larger $M \geq 70$ and/or the smaller  time-step $\Delta t \leq 1/10$.   All of these indicate that the CNS results give the reliable trajectories of the chaotic system, and the CNS is a rigorous approach.          
  
  \begin{table}[htdp]
\caption{Reliable numerical results of  H\'{e}non and Heiles'  chaotic  system (\ref{geq:1}), (\ref{geq:2}) under a different initial condition  (\ref{ic:case2}) given by $ M =  70$ and $\Delta t =1/10$ with  data in accuracy of 140-digit  precision.}
\begin{center}
\begin{tabular}{|c|c|c|} \hline\hline
$t$ & $x(t)$  &  $y(t)$ \\  \hline
500 &  0.19861766  &  -0.23842431  \\
1000 &  -0.04915404 &  -0.31971648  \\
1100 &  -0.48949729  & -0.04052161 \\
1200 &  -0.04886067  &  0.77797896 \\
1300 &   0.42344110  & -0.24151441 \\
1500 &   -0.17190612   &    -0.21349514  \\  
2000 &   0.03364286   &   0.17136302  \\    
\hline\hline
\end{tabular}
\end{center}
\label{Table:case2}
\end{table}%

In addition,  to show the sensitive dependence on initial condition, let us consider a different  initial condition 
\begin{equation}
 x(0)=\frac{14}{25}, y(0)=10^{-60}, \dot{x}(0)=0, \dot{y}(0)=0 \label{ic:case2}
 \end{equation}
with a rather tiny difference of $y(0)$, i.e. $y(0) = 10^{-60}$,  from the previous initial condition (\ref{ic:case1}).   The corresponding CNS results given by $\Delta t =1/10, M=70$ and data in accuracy of 140-digit  precision are listed in Table~\ref{Table:case2}. 
   To verify that it is a reliable trajectory of the chaotic system (\ref{geq:1}), (\ref{geq:2}) and (\ref{ic:case2}) in the interval $0\leq t \leq 2000$,   we repeat the CNS approach by means of $\Delta t = 1/10, 1/20$ and $M = 100, 150, 200, 300, 500$, respectively, and  {\em always}  obtain  the exactly {\em same} results in the interval $t\in[0,2000]$ as those listed in Table~\ref{Table:case2}.  Thus,  the CNS approach indeed provides the true trajectory of the chaotic dynamic system (\ref{geq:1}) (\ref{geq:2}) and (\ref{ic:case2})  in the  restricted  interval $0\leq t \leq 2000$.   
 Note that, the initial condition (\ref{ic:case2}) with $y(0)=10^{-60}$ has a very tiny difference from (\ref{ic:case1}) with $y(0)=0$.    According to Table~\ref{Table:case1} and Table~\ref{Table:case2},  the two reliable  (or shadowing) trajectories  corresponding  respectively  to  the different initial conditions (\ref{ic:case1}) and (\ref{ic:case2}),   match   well  each other  in the interval $0\leq t \leq 1100$.   Even at $t=1200$, they still  match in  accuracy of 5-digit  precision.   However, due to the sensitive dependance  on initial condition,  the two reliable (or shadowing) trajectories  completely  depart  from each other thereafter, although their initial conditions have only a tiny difference in the micro-level $10^{-60}$.  
   
All of these  indicate  that  the CNS results given by $\Delta t =1/10$, $M=70$ and data in the accuracy of 140-digit  precision  are indeed reliable in the interval $t\in[0,2000]$.  In other words, the CNS results given by $M=70$ and $\Delta t =  1/10$  can be regarded as  a kind of  ``shadowing trajectory'' of the chaotic system,  as mentioned by  Dawson et al \cite{Dawson1994}, but in a  restricted interval $0\leq t \leq 2000$.    

It should be emphasized that the difference $10^{-60}$ is indeed rather small, which is however much larger than the truncation error in the level of $10^{-100}$ and the round-off error  in the level of $10^{-140}$  of the CNS approach.   Due to this reason,  the CNS provides us a tool to accurately  investigate the propagation of the micro-level inherent physical uncertainty of chaotic  H\'{e}non-Heiles  system,  which is at the level of $10^{-60}$ that is  much larger than the numerical noises of the CNS, as shown below.        
  
 \section{The micro-level physical uncertainty}
 
 Many, although not all,   mathematical  models  have clear physical background.  A good model for physical problems  often remains the fundamental properties  and  provides  us  a  way to  investigate and predict some of related physical phenomenon.  For example, the law of Newtonian gravitation can describe and predict the motion  of the moon or a satellite accurately.  Besides,  many CFD (computational fluid dynamics) software based on mathematical models can  predict  the  flows  about  a ship and an airplane in an acceptable accuracy.    So, many of mathematical models  reveal   physical  truths of the related  phenomenon.      
 
Eqs. (\ref{geq:1}) and (\ref{geq:2})  provide us  a model for the motion of a star orbiting in a plane about the galactic center, which has very clear physical background.   In general, a good mathematical model  should remain the key physical characteristics  of the corresponding natural phenomena.     Since  the  H\'{e}non-Heiles  system  has been widely accepted by scientific community,  we have many reasons to  believe   that  (\ref{geq:1}) and (\ref{geq:2}) as a mathematical model process the fundamental  physical characteristics of the motion of a star orbiting in a plane about the galactic center.       
 
The kinetic status of a star is determined by its position and velocity.   In the frame of Newtonian gravity law,  it is believed that the kinetic status of a star is {\em inherently} exact and the uncertainty of position and velocity come from the imperfect measure equipments which provide  limited knowledge.    However,  according to de Broglie \cite{Broglie1924},   this traditional idea is wrong:  the position  of a star contains {\em inherent} uncertainty.    Besides,   the quantum fluctuation might influence the  existence of  the so-called  ``objective randomness'',  which is independent of any  experimental accuracy of the observations or limited knowledge of initial conditions,  as suggested  by  Consoli et al \cite{Consoli2011}.     Furthermore,     ``all the sources of complexity examined so far are actually channels for the amplification of naturally occurring randomness in the physical world'', as suggested  by  Allegrini et al \cite{Allegrini2004}.

It is a common belief of the scientific community  that the microscopic  phenomenon are essentially uncertain and random.  To show this point, let us consider some typical length scales of microscopic  phenomenon widely used in modern physics.    For example, 
Bohr radius 
\[ r = \frac{\hbar^2}{m_e \; e^2} \approx 5.2917720859(36) \times 10^{-11} \;\; \mbox{(m)}\]
 is the approximate size of a hydrogen atom, where $\hbar$ is a reduced Planck's constant, $m_e$ is the electron mass, and $e$ is the elementary charge, respectively.   Besides,  
  Compton wavelength  
$ L_c = {\hbar}/{(m c)}   $ 
is a quantum mechanical property of a particle, i.e.  the wavelength of a photon whose energy is the same as the rest-mass energy of the particle, where $m$ is the rest-mass of the particle and $c$ is the speed of light.  It  is the length scale at which quantum field theory becomes important.    The value for the Compton wavelength of the electron is 
\[  L_c \approx 2.4263102175(33) \times 10^{-12}\;\;  \mbox{(m)}.\] 
 In addition,  the Planck length 
\begin{equation}
l_P = \sqrt{\frac{\hbar \; G}{c^3}} \approx 1.616252(81) \times 10^{-35} \;\;\; \mbox{(m)}
\end{equation}
is the length scale at which quantum mechanics, gravity and relativity all interact very strongly,  where  $c$ is the speed of light in a vacuum, $G$ is the gravitational constant, and $\hbar$ is the reduced Planck constant.  Especially, according to the string theory \cite{Polchinshi1998}, the Planck length is the order of magnitude of the oscillating strings that form elementary particles, and {\em shorter length do not make physical senses}.   Besides,  in some forms of quantum gravity, it becomes  {\em impossible}  to  {\em determine} the difference between two locations less than one Planck length apart.   Therefore,  in the  accuracy  of the Planck length level,  the position of a star is  inherently uncertain,   so is its velocity.   Note that this kind of microscopic  physical  uncertainty is {\em inherent} and has nothing to do with the Heisenberg uncertainty principle \cite{Heisenberg1927}  and the ability of human being.

On the other hand,  according to de Broglie \cite{Broglie1924},  any  a  body has the so-called  wave-particle duality, and the length of the so-called de Broglie wave is given by
\begin{equation}
\lambda = \frac{h}{m v}\sqrt{1-\left( \frac{v}{c}\right)^2},
\end{equation}        
where $m$ is the rest mass, $v$ denotes the velocity of the body, $c$ is the speed of light, $h$ is the Planck's constant, respectively. 
Note that,  the de Broglie's wave of a body has non-zero amplitude, meaning that the position is uncertain:  it could be almost {\em anywhere}   along the wave packet.  Thus,  according to  the de Broglie's wave-particle duality,  the position of a star is {\em inherent} uncertain, too.  

Therefore,  it is reasonable  for us  to  assume that  the micro-level inherent fluctuation of position of a star  shorter than the  Planck length $l_p$ is essentially uncertain and/or  random.   
         
To gain the dimensionless  Planck length $l_p$, we use the dimeter of Milky Way Galaxy as the characteristic length,  say,  $d_M  \approx 10^5$ (light year) $\approx 9 \times 10^{20}$ (m).  Obviously,  $l_p/d_M \approx 1.8 \times 10^{-56} $ is a rather small dimensionless number.   As mentioned above,  two (dimensionless) positions shorter than $10^{-56}$ do not make physical senses.  Thus,  it is reasonable to assume the existence of the inherent  uncertainty of the dimensionless position and velocity of a star in the normal distribution with zero mean and  the micro-level standard deviation $10^{-60}$.  Strictly speaking,  such kind of micro-level inherent physical uncertainty should be    added to the {\em observed} values $(x_0, y_0, u_0, v_0)$ of the initial conditions, especially for chaotic dynamic systems whose solutions are rather sensitive to initial conditions.

Therefore, strictly speaking,  the initial condition should be expressed as follows
\[  x(0) = x_0 + \tilde{x}_0,  \;\;  y(0)=y_0 + \tilde{y}_0,  \;\;  
\dot{x}(0) = u_0 +  \tilde{u}_0,   \;\;  \dot{y}(0) = v_0 + \tilde{v}_0,\] 
where $x_0,y_0,u_0,v_0$ are observed values of the initial position and velocity of a star orbiting in a plane about the galactic center, and $\tilde{x}_0, \tilde{y}_0, \tilde{u}_0, \tilde{v}_0$ are the corresponding  {\em micro-level} inherent uncertain ones, respectively.    Assume that $(x_0,y_0,u_0,v_0)$ is exactly given  and  the inherent  uncertain term $(\tilde{x}_0, \tilde{y}_0, \tilde{u}_0, \tilde{v}_0)$  is  in the normal distribution  with zero mean and a micro-level deviation $\sigma =10^{-60}$.   The reasons for the above assumptions  are described above.

Compared to the scale of the initial data $x_0=14/25$,  the deviation  $10^{-60}$ is indeed rather small.  
However,  by means of the CNS approach with the 70th-order Taylor expansion and the MP data in accuracy of 140-digit precision,  the propagation of such kind of micro-level uncertainty can be accurately studied,  because  the corresponding truncation error (in the level of $10^{-100}$) and round-off error (in the level of $10^{-140}$) of the CNS approach is much smaller than the micro-level uncertainty (in the level of $10^{-60}$), as  verified  in \S 2.              

 \section{Statistic property of chaos}
 
Without loss of generality, let us consider the case of the {\em observed} values  \[ x_0=\frac{14}{25}, y_0=0, u_0=0, v_0=0\]  of the initial conditions, corresponding to a chaotic motion \cite{Smith1998}.   The so-called {\em observed} values can be regarded as the mean of measured data.    Let \[  \left<x(t)\right> = \frac{1}{N}\sum_{i=1}^{N}x_i(t),  \hspace{0.75cm}   
 \sigma_x(t) = \sqrt{\frac{1}{N-1} \sum_{i=1}^{N} \left[ x_i(t)-\left<x(t)\right>\right]^2 }\] denote the sample mean and  unbiased estimate of  standard deviation of $x(t)$, respectively, where $N = 10^4$ is the number of total samples,   $x_i(t)$ is the $i$th sample given by the CNS using $\Delta t = 1/10,  M=70$ with the MP data in accuracy of 140-digit precision, and a tiny random term $(\tilde{x}_0,\tilde{y}_0, \tilde{u}_0, \tilde{v}_0)$ with the micro-level deviation $\sigma = 10^{-60}$ in the initial condition.    
 
  \begin{figure}[thbp]
\centering
\includegraphics[scale=0.4]{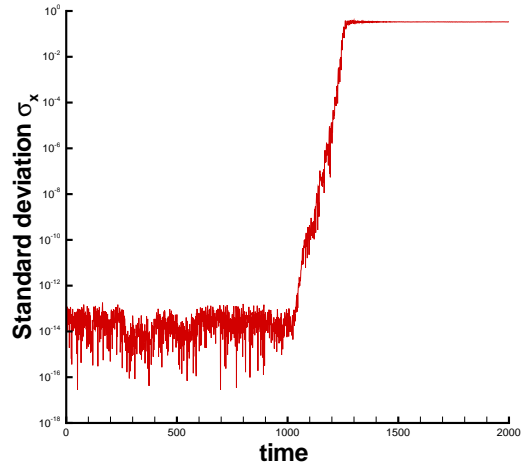}
\caption{The standard deviation $\sigma_x$ of $x$ in case of  $x_0=14/25, y_0=0, u_0=0, v_0=0$ and the uncertain term $(\tilde{x}_0, \tilde{y}_0, \tilde{u}_0, \tilde{v}_0)$  in the normal distribution  with zero mean and a micro-level deviation $\sigma =10^{-60}$. }
\label{figure:sigmaX}
\end{figure}

 \begin{figure}[thbp]
\centering
\includegraphics[scale=0.4]{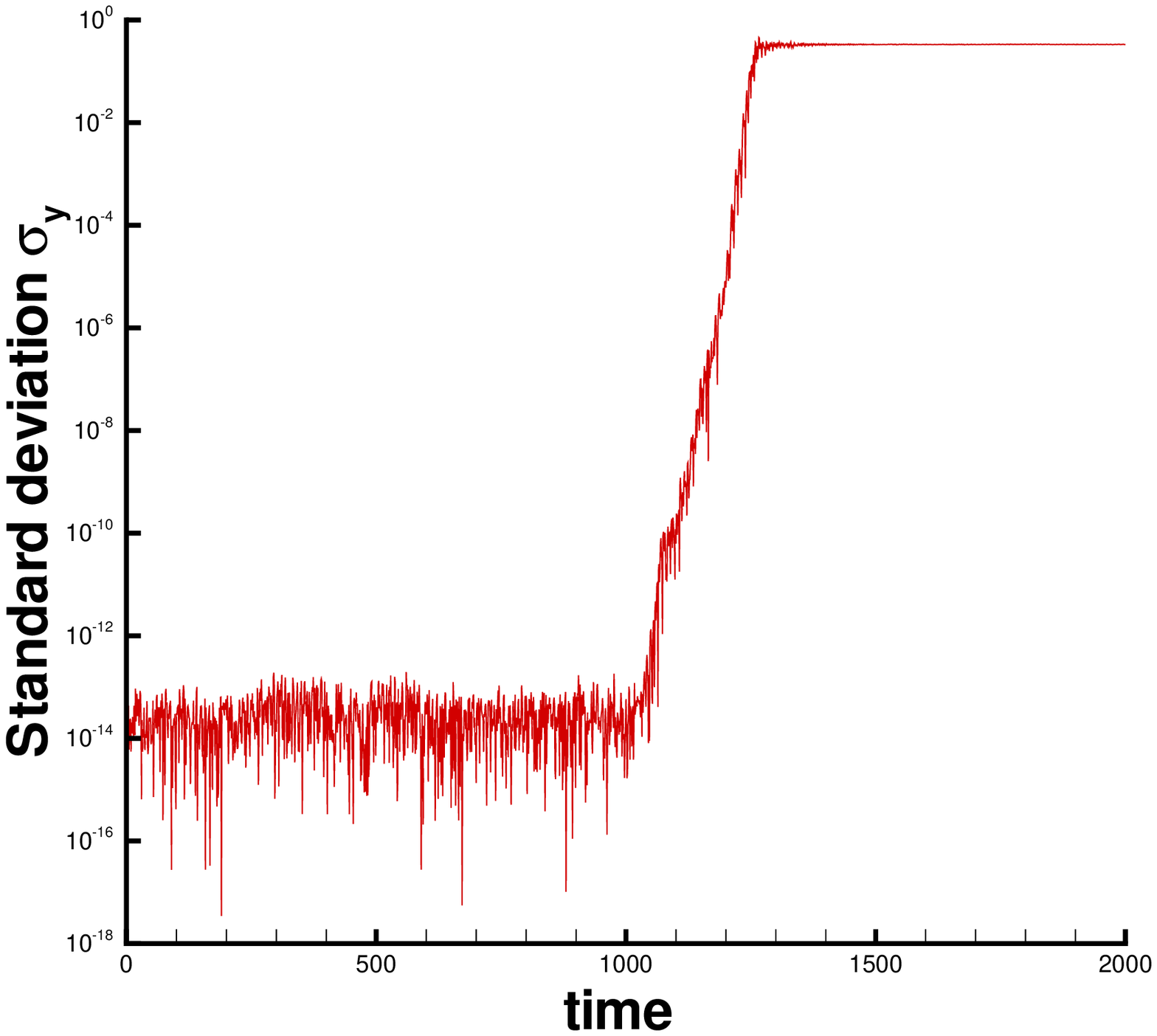}
\caption{The standard deviation $\sigma_y$ of $y$ in case of  $x_0=14/25, y_0=0, u_0=0, v_0=0$ and the uncertain term $(\tilde{x}_0, \tilde{y}_0, \tilde{u}_0, \tilde{v}_0)$  in the normal distribution  with zero mean and a micro-level deviation $\sigma =10^{-60}$. }
\label{figure:sigmaY}
\end{figure}

  \begin{figure}[thbp]
\centering
\includegraphics[scale=0.4]{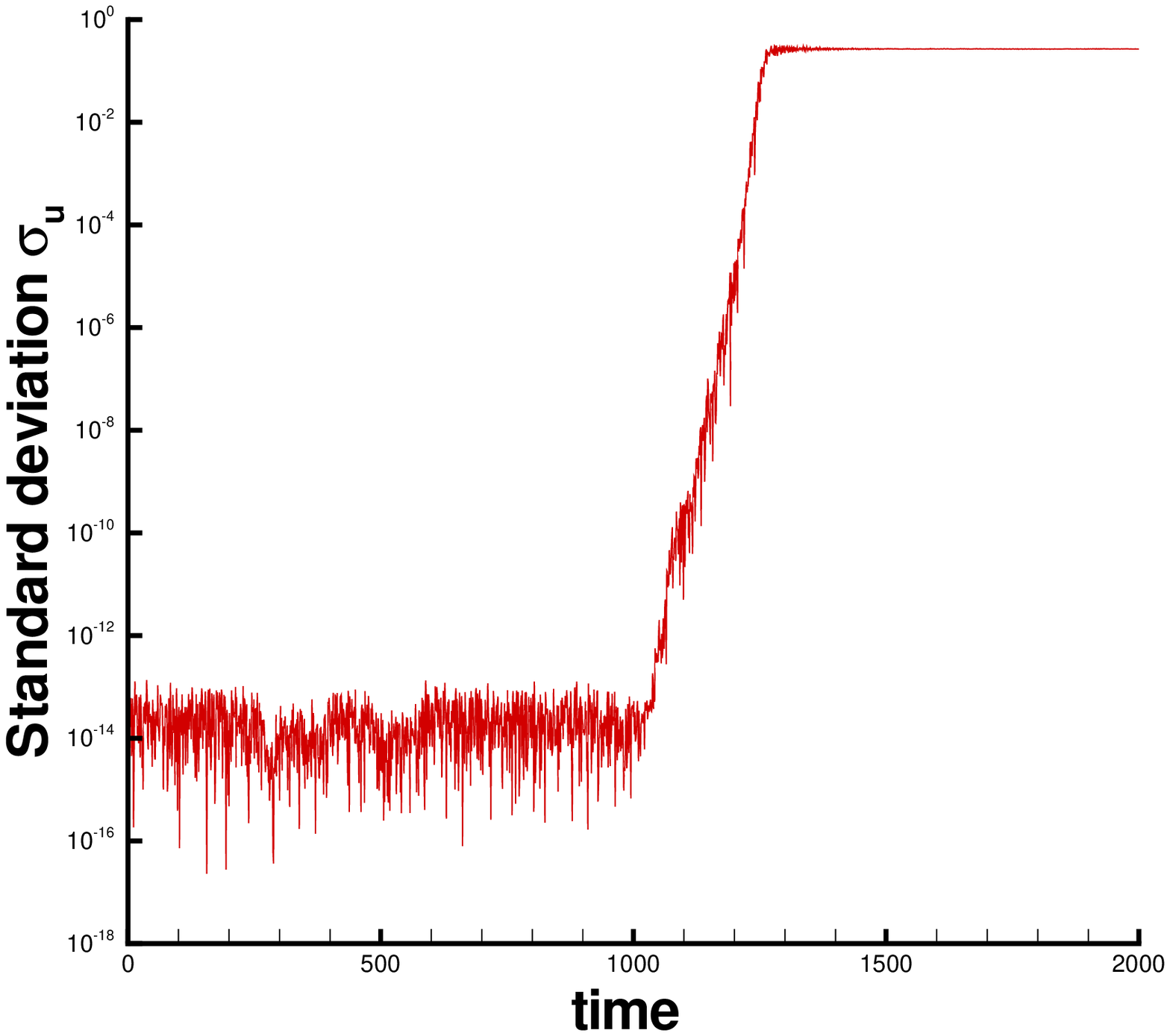}
\caption{The standard deviation $\sigma_u$ of $\dot{x}$ in case of  $x_0=14/25, y_0=0, u_0=0, v_0=0$ and the uncertain term $(\tilde{x}_0, \tilde{y}_0, \tilde{u}_0, \tilde{v}_0)$  in the normal distribution  with zero mean and a micro-level deviation $\sigma =10^{-60}$. }
\label{figure:sigmaU}
\end{figure}

   \begin{figure}[thbp]
\centering
\includegraphics[scale=0.4]{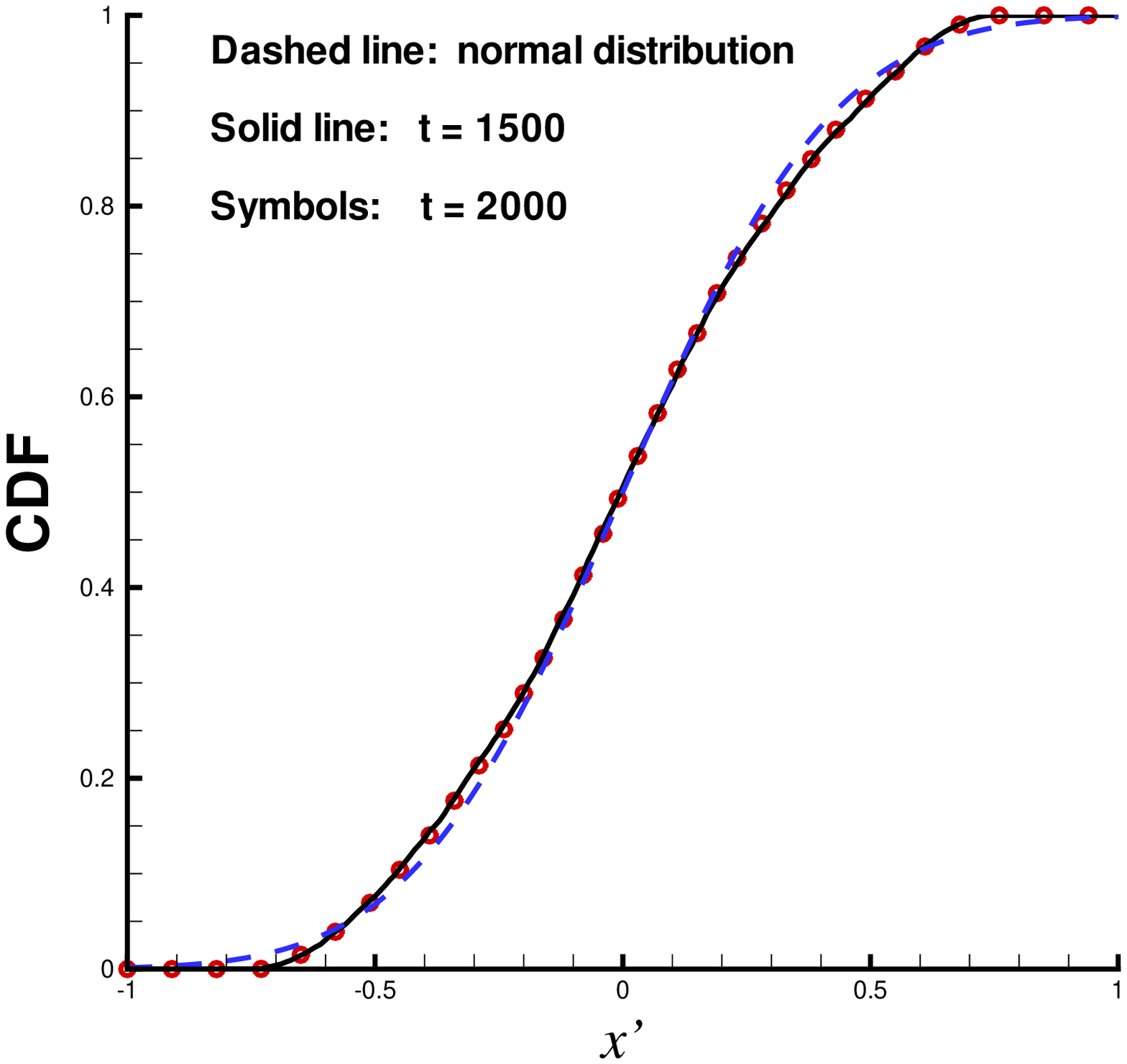}
\caption{The CDF of $x'$, compared to the normal distribution (dashed line)  with zero mean and the standard deviation of $x'$ at $t=2000$.  Solid line: CDF of $x'$ at $t=1500$; symbols:  CDF of $x'$ at $t=2000$. }
\label{figure:CDF-deltaX}
\end{figure}

  \begin{figure}[thbp]
\centering
\includegraphics[scale=0.4]{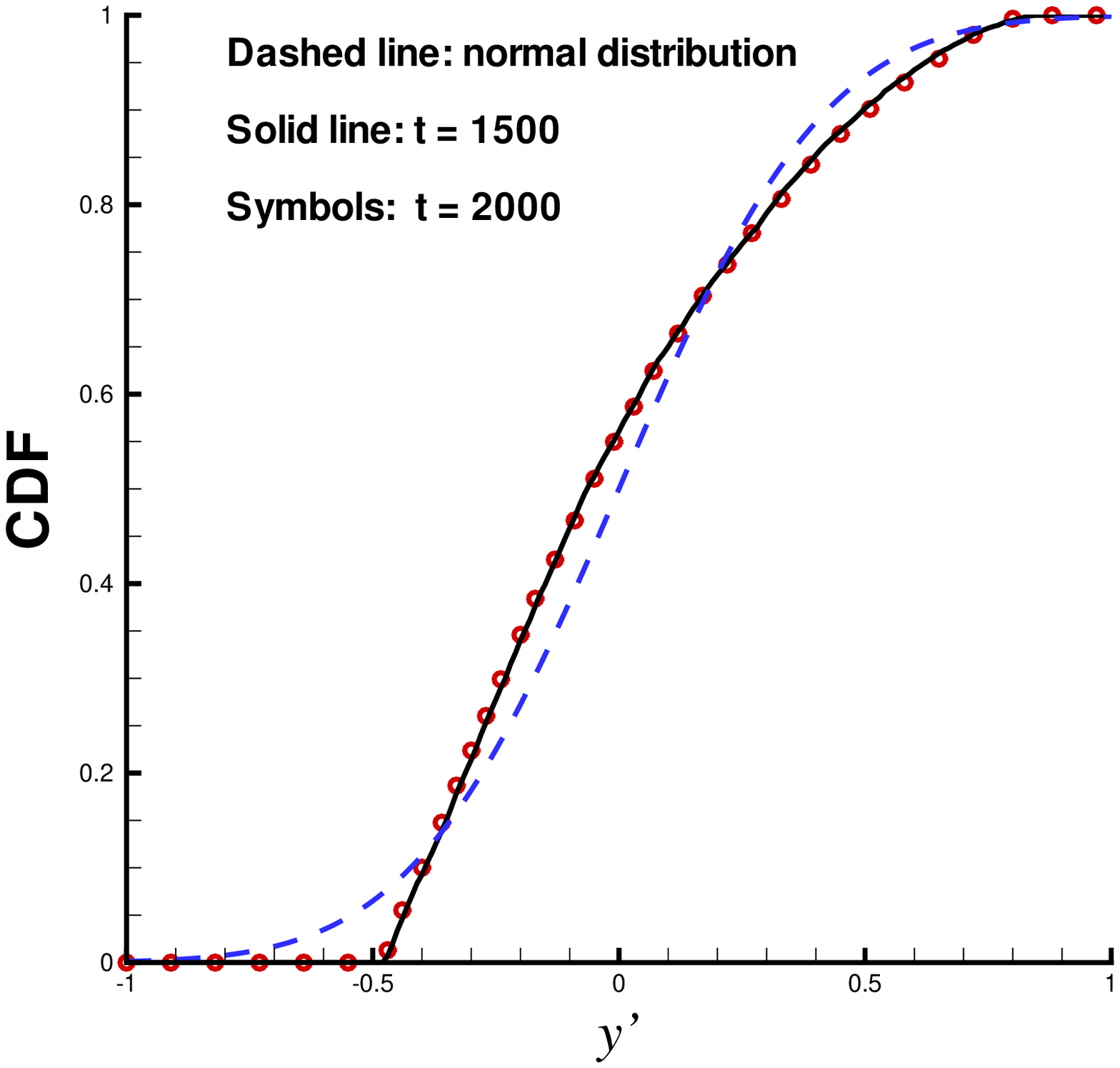}
\caption{The CDF of $y'$, compared to the normal distribution (dashed line) with zero mean and  the standard deviation of $y'$ at $t=2000$.    Solid line: CDF of $y'$ at $t=1500$; symbols:  CDF of $y'$ at $t=2000$.  }
\label{figure:CDF-deltaY}
\end{figure} 

According to \S 2,    all  of  these  $10^4$ trajectories given by the CNS approach  are  reliable  in the interval $t\in[0,2000]$.     The standard deviations $\sigma_x(t)$ and $\sigma_y(t)$ of $x(t), y(t)$ are as shown in Figs.~\ref{figure:sigmaX} and \ref{figure:sigmaY}, respectively.  Note that there exists an interval $0\leq t \leq T_d$ with $T_d\approx 1000$,  in which $\sigma_x(t)$ and $\sigma_y(t)$ are in the level of $10^{-14}$ so that one can accurately predict the position $(x,y)$ of a star, even if the corresponding motion is chaotic  and the initial condition contains uncertainty.   Similarly,  the velocity of the star can be also precisely predicted in $0\leq t \leq T_d$, as shown in Fig.~\ref{figure:sigmaU} for the standard deviation  $\sigma_u(t)$  of $\dot{x}(t)$.   Thus, when $0\leq t \leq T_d$,    the behavior of the chaotic  system looks like ``deterministic'' and ``predictable'', even from the statistic viewpoint.     When $t>T_d$,   the standard deviations of the position and velocity begin to increase rapidly, and thus the system becomes  random obviously:  the position $(x,y)$ and velocity $(\dot{x},\dot{y})$ of the star are strongly dependent upon their micro-level  inherent  physical uncertainty  $(\tilde{x}_0, \tilde{y}_0, \tilde{u}_0,\tilde{v}_0)$ of the initial condition.   In other words,  due to the SDIC of chaos,  the unobservable  micro-level inherent uncertainty of the position and velocity of a star transfers into the macroscopic randomness of the motion.    This suggests that  chaos  might be a bridge from the micro-level uncertainty  to macroscopic randomness!   Therefore,  the micro-level inherent uncertainty of the position and velocity might be an origin of the macroscopic randomness of motion of stars in our universe.    Possibly,  this  might provide us a new,  {\em physical} explanation and understanding for the SDIC of chaos.   For this reason,  each ``big bang''  \cite{Peacock1998}  will crease a completely different universe!     

Besides,   it is found that the standard deviations of the position and velocity   become  almost stationary when  $t  > T_s$,  where $T_s \approx 1300$,  as shown in Figs.~\ref{figure:sigmaX} to \ref{figure:sigmaU}.   Thus,  when $T_d < t < T_s$,  the system is in the transition process from the ``deterministic'' behavior to the stationary randomness.   It is interesting that the stationary  standard deviations of $x(t)$ and $y(t)$  are  about 1/3, and their stationary means $<x>$ and $<y>$  are close to zero.     It means that, due to SDIC of chaos  and   the micro-level inherent uncertainty of position and velocity,   a  star orbiting in a plane about the galactic center could be almost {\em everywhere} in the galaxy at a given time $t > T_s$.    
 
Write the fluctuations $x'(t) = x-<x>$ and $y'(t) = y-<y>$.    The stationary cumulative distribution functions (CDF)  of $x', y'$  are almost independent of time, as shown in Figs.~\ref{figure:CDF-deltaX} and \ref{figure:CDF-deltaY}.   Besides, the stationary CDF of the fluctuation $x'$  is rather close to the normal distribution with zero mean and the standard deviation of $x'$, as shown in Fig.~\ref{figure:CDF-deltaX}.  But, the stationary CDF of the fluctuation $y'$ is obviously different from the normal distribution, as shown in Fig.~\ref{figure:CDF-deltaY}.  

Similarly, we investigate the influence of the observed values $(x_0, y_0, u_0, v_0)$ and the standard deviation  $\sigma$ of the uncertain terms $(\tilde{x}_0, \tilde{y}_0, \tilde{u}_0, \tilde{v}_0)$  in the initial condition by means of the CNS approach.    It is found that $T_d$ decreases exponentially with respect to $\sigma$.  Besides,  the stationary means and standard deviations of $x, y, \dot{x}, \dot{y}$, and the   CDFs of $x'$ and $y'$,  are independent of the observed values  $(x_0, y_0, u_0, v_0)$.   Thus,  when $t > T_s$,   all observed  information of the initial condition are lost completely.   In other words,   when $t  > T_s$,  the asymmetry of time breaks down so that the time has a one-way direction, i.e.  the arrow of time.   So, statistically speaking,  the H\'{e}non-Heiles system has two completely different dynamic behaviors before and after $T_d$:  it looks  like  ``deterministic'' and ``predictable''  without time's arrow when $t\leq T_d$, but thereafter rapidly becomes  obviously  random with a arrow of time. 

 Consoli et al \cite{Consoli2011} suggested that the objective randomness  ``might introduce a weak, residual form of noise  which is  intrinsic to natural phenomena and could be important  for the emergence of complexity at higher physical levels''.   Our extremely accurate numerical simulations  based on the CNS approach  support their  viewpoint:  the micro-level uncertainty and  the macroscopic randomness   might have a rather close relationship.  

\section{Conclusions and discussions}

In this paper, an extremely accurate  numerical algorithm, namely the ``clean numerical simulation'' (CNS), is proposed to accurately simulate the propagation of micro-level inherent physical  uncertainty of chaotic dynamic systems.   The chaotic  H\'{e}non-Heiles system describing the motion of a star orbiting in a plane about the galactic center  is used as an example to show the validity of the CNS approach.   

In the frame of the CNS approach,    the truncation error is estimated by (\ref{bounded}),  the  round-off error is determined by the digit-length of data, and the critical time $T_c$ is explicitly  determined  by (\ref{Tc:M}) (for the chaotic H\'{e}non-Heiles system).   So, given an arbitrary value of  $T_c$,  we can always find out the required order $M$ of Taylor expansion  and the data in accuracy of $2M$-digit precision so as  to gain a reliable trajectory of the chaotic H\'{e}non-Heiles   system in the finite interval $t\in[0, T_c]$  by means of $\Delta t =  1/10$.   In addition,   the CNS results in the interval $t\in[0,T_c]$ are verified  very carefully by means of Taylor expansion at higher-order and MP data in more accuracy, as shown in \S 2.     As shown in \S 2,   {\em all}  of the  CNS results (for the same initial condition) given by $\Delta t =1/10, M=70$ and data in accuracy of 140-digit precision are exactly the {\em same} as those given by  $M=100, 150, 200, 300, 500$ and $\Delta t =1/20, 1/100$, respectively, so that they are indeed reliable, true  trajectories of the chaotic H\'{e}non-Heiles system.   Besides,   as the order $M$ of Taylor expansion increases and the time-step $\Delta t$ decreases,  the truncation and round-off errors decrease  monotonously.  For example,  
as illustrated in Tables~\ref{Table:error:dt=0.1} to \ref{Table:error:dt=0.01},  the truncation error  is in the level of $10^{-100}$  in case of $\Delta t = 1/10$ and $M=70$, and decreases to the level $10^{-1244}$ in case of $\Delta t=1/100$ and $M=500$.   In addition, the round-off error is simply in the level of $10^{-2M}$, where $M$ denotes the order of Taylor expansion.    So, theoretically speaking, one can control the truncation and round-off error to a required level.   In these meanings, the CNS approach is a rigorous one.  

The H\'{e}non-Heiles system of (\ref{geq:1}) and (\ref{geq:2}) as a mathematical model  has clear  physical background:    it  has been widely accepted and used by the scientific community to  describe  the motion of a star orbiting in a plane about the galactic center.   The status of a star is dependent  upon its position and velocity.    However,  according to de Broglie \cite{Broglie1924},   the position  of a star contains micro-level  {\em inherent}  physical  uncertainty, as discussed in \S 3.   So, strictly speaking,  the H\'{e}non-Heiles system of (\ref{geq:1}) and (\ref{geq:2}) is {\em not} deterministic in essence.   Due to the SDIC of chaos,  such kind of micro-level physical uncertainty transfers into macroscopic randomness of motion, as  illustrated  in \S 4 by means of the CNS approach.    Therefore,  the micro-level inherent physical uncertainty and macroscopic randomness might have a close relationship:   chaos might be a bridge from the micro-level inherent physical uncertainty to macroscopic randomness!  This conclusion  agrees with the viewpoint of Consoli et al \cite{Consoli2011} who suggested that the objective randomness  ``might introduce a weak, residual form of noise  which is  intrinsic to natural phenomena and could be important  for the emergence of complexity at higher physical levels''.

The CNS approach provides us an extremely precise numerical approach for chaotic dynamic systems in a given {\em finite} interval $t\in[0,T_c]$.  According to (\ref{Tc:M}), $T_c\to +\infty$ as $M\to +\infty$.  In other words,  if the initial condition were exact,  then long-term prediction of chaos  would be  possible  {\em in theory} \footnote{Unfortunately,  the required  CPU time increases exponentially as $M$ increases,  so that it is {\em practically} impossible to give reliable, true trajectories of chaos in a very large interval.} :  given an arbitrary value of  $T_c$,    we  could  gain the reliable chaotic trajectory of the  H\'{e}non-Heiles system in the {\em finite} interval $0\leq t \leq T_c$ by means of the $M$th-order Taylor expansion with data in accuracy of $2 M$-dight precision, as long as  $M > T_c/32$ and $\Delta t \leq 1/10$.   Qualitatively,  the  conclusion has general meanings and holds for other chaotic models such as Lorenz equation.     Besides, it  is  consistent   with   Tucker's  elegant  proof \cite{TuRODES, Tu} that there indeed exists an attractor of Lorenz equation.    Thus,  theoretically speaking,  there is no place for the randomness in a {\em truly} deterministic system.   However,  most models related to physical problems contain more or less physical uncertainty, and thus, strictly speaking,  are {\em not}  deterministic.  For such kind of physical models with inherent uncertainty, the accurate long-term prediction of trajectories of chaotic system has {\em no} physical meanings, because their long-term trajectories are {\em inherently} random that comes from the micro-level inherent physical uncertainty, as  illustrated  in this article.  

Traditionally, it is believed  that  the  long-term prediction of chaos is impossible,  mainly due to the impossibility of the  perfect  {\em  measurement}  of initial conditions with an {\em arbitrary} degree  of accuracy.   This  is  the traditional explanation to the SDIC of chaos.     Here,  we  provide a new explanation for the SDIC of chaos from the physical viewpoint:  initial conditions of some chaotic systems with clear  physical meanings  might contain  micro-level  {\em inherent}  physical  uncertainty, which   might   propagate  into macroscopic randomness. 
Different from the traditional explanation of the SDIC, which focuses on the {\em measurement},  the new explanation emphasizes the {\em inherent} micro-level uncertainty  and its propagation with chaos.   Besides,  it should be  emphasized  that  such micro-level inherent physical uncertainty of chaos  was  completely  inundated  with  the numerical noises of the traditional numerical methods based on low-order algorithms,  and thus has never been  studied  in details.    This  shows the validity and potential of the CNS to precisely simulate complex physical phenomena with the SDIC,  such as weather prediction and   turbulence.
 
 \begin{table}[t]
\caption{Mapping of $f(x) = \mbox{mod}(2x,1)$ with $x_0=\pi/4$, expressed in decimal and binary systems}
\begin{center}
\begin{tabular}{|c|c|l|} \hline\hline
$n$  &  $x_n$ in decimal system    &  \hspace{1.25cm} $x_n$  in binary system \hspace{1.0cm}  \\ \hline
0  &  0.785398163397448 $\cdots$   &  110010010000111111011010 $\cdots$ \\
1&    0.570796326794896 $\cdots$   &  10010010000111111011010 $\cdots$\\
2&    0.141592653589793 $\cdots$   &  10010000111111011010 $\cdots$\\
3&    0.283185307179586 $\cdots$   &  10000111111011010 $\cdots$\\
\vdots &  \vdots & \vdots \hspace{2.5cm} \\
\hline\hline
\end{tabular}
\end{center}
\label{Table:map}
\end{table}%

Finally, for the easier understanding of the CNS, let us consider the map 
\begin{equation}
f(x) = \mbox{mod} (2x,1)  \label{map:mod}
\end{equation}   
with the initial value $x_0 = \pi/4$.  It is well-known that this map has the sensitivity dependence on initial condition, i.e. SDIC.  The results of the $n$th mapping, i.e. $x_n = f(x_{n-1})$ with $x_0=\pi/4$,  are  expressed   by both of the decimal and binary systems in Table~\ref{Table:map}.  In binary system,  the mapping $x_n$ corresponds to such a kind of left shift:  shifting $x_0$  left to the position of its 2nd digit ``1'' gives $x_1$, and to the position of its 3rd digit ``1''  gives  $x_2$, and so on, as shown in Table~\ref{Table:map}.  In general, $x_n$ (in binary system) corresponds to the left shift of $x_0$  (in binary system) to its position of the $(n+1)$th digit ``1''.    Since $\pi/4$ is exactly known in binary system, its position of the $n$th digit ``1'' is deterministic, denoted by $P_2(n)$.   So, in binary system,  $x_n$ is exactly the left shift of $x_0$ to its  $P_2(n+1)$-th digit  ``1''.   So, mathematically speaking,  this mapping is deterministic and $x_n$ is exactly known.     However,  in practice,  one had to  take $x_0 = \pi/4$ in a {\em finite} accuracy, which leads to uncertainty.  Assume that $x_0$ is in accuracy of $N_0$ binary digits.  Then,  $x_1, x_2,x_3$ has $N_0-1, N_0-4, N_0-7$ significance binary digits,  respectively, as shown in Table~\ref{Table:map}.     In general,  $x_n$ is in the accuracy of $N_0 - P_2(n+1)$  binary  digit  precision.    Obviously,  when $P_2(n+1) > N_0$, the mapping $x_n$ losses its accuracy at all.  However,  even at one million times of mapping, i.e. $n=10^6$,  we can gain the {\em accurate enough} result $x_{1000000}$ in the accuracy of one million of binary digit precision, as long as we take the initial value $x_0 = \pi/4$ in the accuracy of $10^6 + P_2(10^6+1)$ binary digit precision!   This simple example illustrates that we {\em do} can gain reliable results for dynamic systems with SDIC in a {\em finite} times of mapping or a {\em finite} interval, as long as initial  conditions are  {\em accurate enough}.   This also explains why the CNS is based on rather accurate data,  using the computer algebra system Mathematica or the multiple precision library.                             

\begin{table}[t]
\caption{$x_n$ given by the mapping $f(x)=\mbox{mod} (2x,1)$ with the initial value $x_0=\pi/4$ in different accuracy of $N$ decimal digit precision. }
\begin{center}
\begin{tabular}{|c|l|l|l|l|l|l|}\hline\hline
$n$  & $\hspace{0.5cm} N = 15$	& \hspace{0.5cm} $N  = 20$	& \hspace{0.5cm} $N = 25$ 	& \hspace{0.5cm}	$N = 30$ 	& \hspace{0.2cm} $N = 1000$  \\ \hline
5	&	0.1327412287	&	0.1327412287	&	0.1327412287 	& 	0.1327412287  	& 	0.1327412287	\\
10	&	0.2477193189	&	0.2477193189 	&	0.2477193189	&	0.2477193189		&	0.2477193189	\\
15	&	0.9270182076	&	0.9270182075 	&	0.9270182075	&	0.9270182075		&	0.9270182075	\\
20	&	0.664582643 	&	0.6645826427 	&	0.6645826427	&	0.6645826427		&	0.6645826427	\\
25	&	0.2666446		&	0.2666445682	&	0.2666445682	&	0.2666445682		&	0.2666445682 	\\
30	&    0.532626		&	0.5326261849 	&	0.5326261849	&	0.5326261849		&	0.5326261849	\\
35	&	0.0440			&	0.044037917  	&	0.0440379171	&	0.0440379171		&	0.0440379171	\\
40	&	0.409 			&	0.40921335   	&	0.4092133503	&	0.4092133503 		& 	0.4092133503	\\
\hline\hline
\end{tabular}
\end{center}
\label{Table:map:2}
\end{table}%

However,  a chaotic dynamic system has no such kind of elegant property of $\mbox{mod}(2x,1)$ mentioned above, since its  exact solution is   unknown  in general.  Thus,  the above approach based on the left shift has no general meanings.   Assume that one  knows   the SDIC of the mapping $f(x) = \mbox{mod}(2x,1)$,  but  has  no  ideas  about  its  left-shift  property  in the corresponding binary system.   How to gain reliable sequence $x_n = f(x_{n-1})$  by means of  $x_0=\pi/4$?   A  general, straight-forward way  is to compare two sequences given by $x_0=\pi/4$ in different accuracy of $N$-digit precision (in decimal system), where $N=15, 20, 25, 30$ and 1000, respectively, as shown in Table~\ref{Table:map:2}.        For example,  by comparing the two sequences of  $x_n$ given by $x_0$ in  accuracy of 15 and 20-digit precision,   one is sure due to the SDIC that  the sequence $x_n$  given by $x_0$ in accuracy of 15-digit precision  is  reliable  at $n\leq 15$ in accuracy of  8 significance digits.    Similarly,  using $x_0$ in accuracy of 20 and 25-digit precisions,   one gains reliable $x_n$  at $n\leq 30$ and $n\leq 40$ with 8 significance digits, respectively.    Note that the sequence $x_n$ given by $x_0$ in accuracy of 25-digit precision agrees well with that by $x_0$ in accuracy of 30-digit precision for a {\em finite} number of mappings $x_n$, where  $0\leq n \leq 40$.   Thus,   one has many reasons to believe that  the {\em finite} sequence $ x_0, x_1, \cdots, x_{40}$  given  by means of $x_0$ in accuracy of 25-digit precision is reliable.   This is indeed true,  because it  completely agrees with the  ``exact''  sequence given by $x_0 $ in accuracy of 1000-digit precision, as shown in Table~\ref{Table:map:2}.   The  key  point  is that, to gain reliable sequence  $x_0, x_1, \cdots, x_{40}$ with the {\em finite}  number of mappings,  we need use $x_0$ in  accuracy  of  {\em only} 25-digit precision:  it is unnecessary to use $x_0$ in higher accuracy.   Similarly,  using $x_0$ in accuracy of 40-digit precision, one can gain reliable sequence \[ x_0,x_1, x_2, \cdots, x_{100}\]  in accuracy of 8 significance digits.  Furthermore,  using $x_0$ in accuracy of 60-digit precision, one can gain reliable sequence 
\[  x_0, x_1, x_2, \cdots, x_{166}\]
in accuracy of 8-digit precision as well.   And so on.   Thus,  we can gain  reliable $x_n$ of {\em finite but many enough } mappings by using {\em accurate enough} $x_0$.    In other words,  the  reliability  and precision of the {\em finite} sequence \[   x_0, x_1, x_2, \cdots, x_n \] given by the mapping $f(x)=\mbox{mod}(2x, 1)$ with SDIC   is under {\em control}.    It is true that, using $x_0$ in 60-digit precision,  $x_{100000}$ is incorrect and thus has no meaning.   However,  the key point is that the corresponding  sequence  
\[  x_0, x_1, x_2, \cdots, x_{166}  \] 
of a {\em finite} number of mappings  is reliable in accuracy of 8-digit precision,  which might be enough for one's purpose.   In essence,  we seek for a kind of {\em relative} reliability and predictability  of chaotic dynamic systems,   although  very  long-term  accurate  prediction of any a chaotic dynamic system  is  absolutely  impossible in theory.   It is true that, using $x_0$ in accuracy of any a given precision, there {\em absolutely}  exists such a large enough $n$ that  $x_n$ totally  losses  its accuracy.    However,  we can guarantee  the reliability and predictability of a given  {\em finite} sequence (such as $x_0,x_1, x_2, \cdots, x_{166}$) by using $x_0$ in a reasonable accuracy (such as  60-digit precision).  It should be emphasized that  such kind of comparison approach  is valid  for any chaotic dynamic systems.  So,   it has general meanings and thus is practical.   Note that the {\em same}  comparison approach  is used in the CNS  described  in \S 2 and \S3.     This example  clearly  explains why the CNS based on such kind of comparison is indeed reasonable and valid.    

It is important to provide a {\em practical} numerical approach to gain reliable chaotic solutions of dynamic systems in a long enough interval of time.  Using CNS with 400-order Taylor expansion,  data in accuracy of 800-digit precisions and $\Delta t=10^{-2}$,  Liao \cite{Liao2009} gained, for the first time, a  reliable chaotic solution of Lorenz equation in a rather long time interval $0 \leq t \leq 1000$, whose correction is confirmed by Wang et al.  \cite{Wang2011}.      As mentioned by Wang \cite{Wang2012PhD},  in order to gain reliable chaotic solution of Lorenz equation in the interval $0\leq t \leq 1000$ by means of the 4th-order Runge-Kutta method, one had to use  multiple precision data in $10000$-digit precision and a rather small time-step $\Delta t =10^{-170}$,  which however needs about $3.1 \times 10^{160}$ years by today's high-performance computer!  Therefore, the low-order Taylor expansion approaches are {\em not} practical to gain reliable chaotic solution of Lorenz equation in such a long time interval.  There exist some ``rigorous'' simulations \cite{Tu}  assuring  that the real orbits of chaotic system are ``enclosed''  in a computed region of space, such as $[x(t) -\delta, x(t) +\delta]$, where $\delta$ should be a small constant: results with large $\delta$ is useless  in practice, even though it is  obtained by ``rigorous'' methods.   Due to SDIC, it is obvious that one had to use rather small $\delta$ to gain such a rigorous chaotic solution of Lorenz equation in $0\leq t \leq 1000$ by means of the enclosing approach:  possibly $\delta$ might be in the level  of $10^{-480}$, since the corresponding initial condition must be accurate in 480 digit precision, as pointed out by Liao \cite{Liao2009}.  However, to the best of author's  knowledge,  it is still an {\em open} question whether or not the ``rigorous''  method based on enclosing  \cite{Tu}  can give such a reliable, accurate enough chaotic solution of Lorenz equation  in the interval $0\leq t\leq 1000$ by means of a reasonable CPU time.     Besides,  to the best of author's knowledge,  it is also an {\em open} question whether or not the enclosing approach is {\em practical} for  physical problems like those considered in this article: note that the CNS is successfully used to gain 10000 samples of reliable chaotic solutions given by different initial conditions with $10^{-60}$-level uncertainty.   So, compared to other approaches,  the CNS  is  not  only  {\em reliable}  but  also  {\em practical}. 

Indeed,  the propagation of round-off and truncation errors of a chaotic dynamic system is rather complicated and thus is unknown in general cases.  As pointed out by Parker and Chua \cite{Parker1989}, a ``practical''  way of judging the accuracy of numerical results of a non-linear dynamic system is to use at least two (or more) ``different'' routines to integrate the ``same'' system.   This is mainly because, due to the SDIC of chaotic dynamics systems,   departure of two chaotic simulations  indicates  the  appearance  of  large enough  truncation  and  round-off  errors.   In practice, the comparison approach provides us  
a time interval $0 \leq t \leq T_c$, in which  the  same  results  should  be  reliable,  mainly due to SDIC of chaos.   Certainly, such kind of  critical time $T_c$ must be carefully checked by as many different approaches as possible, as shown in \S2.2.       In fact, such kind of comparison approach is widely accepted by scientific community \cite{Parker1989, Teixeira2007, Wang2011, Wang2012PhD}.  And the CNS is based on such kind of strategy.   Using a metaphor,  it is like measuring the height of a man: the better the equipment,  the more accurate the result, although we can not provide an ``exact'' value of the height.   Although it is difficult to measure the height of a man in accuracy of $10^{-10}$ meter,  it is rather easy to ensure that whether  a man is higher than 1.85 meter or not, as long as all measures given by all equipments  give us the same answer to this question:  such kind of precision is relatively rough but enough in many cases of everyday life.  Similarly, the CNS seeks for reliable, accurate enough simulations of chaotic dynamic systems in a {\em finite} time-interval.     

In summary, the CNS provides us a {\em practical} way to gain reliable, {\em  accurate enough}  solutions of chaotic dynamic systems  with a {\em high enough} precision in a {\em finite} but {\em long enough} time interval.            

\section*{Acknowledgement}  

The author would like to express his sincere thanks  to  the two reviewers for  their  comments, which greatly  heighten  the  quality  of  this  article.   Thanks  to Prof.  Y. L. Bei (Chinese Academy of Sciences),  Prof. M.G. Xia (Chinese Academy of Sciences), Prof. H.R. Ma (Shanghai Jiao Tong University) for their valuable discussions.  

%\pagebreak\newpage 

\end{document}